# Insights on the mechanism of water-alcohol separation in multilayer graphene oxide membranes: entropic versus enthalpic factors

*Daiane Damasceno Borges[1], Cristiano F. Woellner[1,2], Pedro A. S. Autreto[3], and Douglas S. Galvao[1]*

[1]Applied Physics Department, University of Campinas - UNICAMP, Campinas-SP 13083-959, Brazil;

[2]Department of Materials Science and Nano Engineering, Rice University, Houston, Texas, USA;

[3]Center for Natural and Human Sciences, Federal University of ABC - UFABC, Santo Andre-SP, 09210-580, Brazil

AUTHOR EMAIL ADDRESS daianefis@gmail.com, galvao@ifi.unicamp.br

**ABSTRACT**

Experimental evidences have shown that graphene oxide (GO) can be impermeable to liquids, vapors and gases, while it allows a fast permeation of water molecules. The understanding of filtration mechanisms came mostly from studies dedicated to water desalination, while very few works have been dedicated to distilling alcohols. In this work, we have investigated the molecular level mechanism underlying the alcohol/water separation inside GO membranes. A series of molecular dynamics and Grand-Canonical Monte Carlo simulations were carried out to probe the ethanol/water and methanol/water separation through GO membranes composed of multiple layered graphene-based sheets with different interlayer distance values and number of oxygen-containing functional groups. Our results show that the size exclusion and membrane affinities are not sufficient to explain the selectivity.

Besides that, the favorable water molecular arrangement inside GO 2D-channels forming a robust H-bond network and the fast water diffusion are crucial for an effective separation mechanism. In other words, the separation phenomenon is not only governed by affinities with the membrane (enthalpic mechanisms) but mainly by the geometry and size factors (entropic mechanisms). We verified that the 2D geometry channel with optimal interlayer distance are key factors for designing more efficient alcohol-water separation membranes. Our findings are consistent with the available experimental data and contribute to clarify important aspects of the separation behavior of confined alcohol/water in GO membranes.

**KEYWORDS**

Graphene oxide, alcohol dehydration, separation membrane, computer simulations

**INTRODUCTION**

Graphene oxide (GO) membranes have been investigated as very promising candidates for water filtration and/or separation membranes[1,2]. Recently, experimental evidences have shown that in the aqueous phase, water molecules can permeate through GO membranes while blocking the passage of ions[3] and molecules such as ethanol[4], methanol, propanol[5] and others[6]. GO membranes can exhibit complex topologies depending on the experimental techniques used to fabricate them. They can lead to distinct micro- and nano- morphologies and transport pathways that directly affect their selectivity properties. The molecules inside the membrane are supposed to be transported through the percolated two-dimensional channels formed between graphene-like sheets. A single GO sheet structure contains high percentage of functionalized regions (~82%), pristine regions (~16%) and structural holes (~2%)[7]. The pristine regions have the same local structure as pristine graphene where all atoms are bonded in $sp^2$ hybridization. The functionalized regions have a large amount of hydroxyl, epoxy and carbonyl groups and the atoms are bonded in $sp^3$-like hybridizations. The number of functional groups can be tunable through some chemical post-treatments. The typical interlayer distance between GO sheets is ~6-7 Å under dry conditions and about 12-13 Å under humidification, and it can be also controlled by physical confinement, as recently shown by experimentalists[8,9]. The tunable structural properties by experimental techniques combined to a better understanding of the structural role on the separation mechanism are crucial for designing efficient selective membranes.

The promising application of GO membranes for water removal from alcohols, such as ethanol and methanol, would have important relevance in biofuels production processes. Ethanol is the most commercialized biofuel that might help world energy supply demands with minor negative environmental impacts[10]. However, to improve sustainability and cost gain of this renewable resource, efficient separation membrane is still needed to reduce energy cost during the process of ethanol dehydration[11]. In order to separate water-alcohol mixture, there is a series of steps that makes the origin of GO membranes selectivity not evident. GO presents the opposite behavior compared to pure graphene that adsorbs more alcohols instead of water.[12] The graphene oxide is much more hydrophilic, making the water rejection improbable. Indeed, both water and alcohol are polar molecules and they would preferentially interact with the polar oxygen-containing functional groups of GO in a similar way. Also, water with methanol and ethanol exhibit comparable molecular sizes. Thus far, the role of GO membrane in the water-alcohol separation remains unclear, limiting further improvements on their selectivity performance.

Theoretical studies devoted to understand the water filtration mechanisms inside GO membranes came mostly from studies dedicated to water desalination[13], while very few works have been dedicated to water-alcohol separation. The separation mechanism has been attributed to the formation of a network of nano-capillaries that allow nearly frictionless water flow, while blocking other molecules by steric effects[4]. In this work, we propose a structural model that mimics a fluid flowing from an infinite reservoir through a multilayered GO-based membrane by molecular dynamics (MD) simulations. This method allows the investigation of the capillary effects that induce the flow, considering the guest diffusion and the wetting process of the membrane. In addition, we have also carried out Grand-Canonical Monte Carlo (GCMC) simulations to probe ethanol/water and methanol/water adsorptive separation. To obtain a deep understanding on these mechanisms at molecular scale, a careful inspection on the structural aspects (how the molecules are spatially arranged) of the molecules inside the membranes, as well as membrane/molecule affinity and molecular diffusion were analyzed.

**RESULTS AND DISCUSSION**

The permeability of alcohols (*i.e.* methanol and ethanol) and water within graphene oxide (GO) membranes were systematically investigated using MD simulations to mimic GO multilayers membranes in contact with a reservoir of water-alcohol mixtures of 0.5 mole fraction. The pure liquid water reservoir was also simulated for comparison purposes. Two types of membranes were built with GO sheets functionalized with hydroxyl and epoxy on both sides and carboxyl groups on the sheet edges, totalizing O/C ratio equal to 35% and 14%, respectively (see Figure 1). At the simulation runs, the membrane induces the fluid to flow into the nanocapillary channels. The fluid pathway is determined by the nanocapillary network formed from connected interlayer spaces (2D channels), together with the slits in the GO sheets. Two interlayer distances were considered here: 7 Å (GO7) and 10 Å (GO10), while the slit width was fixed to be equals to 20 Å, as indicated in Figure 1. The flow is driven by the difference of pressure between the inlet of the membrane (at the bottom), which is the water/alcohol reservoir and the outlet of the membrane (on top), which is vacuum. The difference of pressure was controlled by the piston on the bottom of the reservoir, in order to keep the same value for all considered cases.

The water/alcohol separation occurs right after the fluid starts flowing upwards into the membrane. The favorable permeation of water against alcohols is evident in all investigated cases, pointing out that the results of our model are consistent with the separation phenomena observed in several experiments[14,15]. Figure 2 shows the flux differences of water *vs*. ethanol and water *vs*. methanol inside GO7 (35%). The graphs display the time evolution of the number of water and alcohol molecules flowing outside the reservoir into the membrane. The membrane becomes filled with water rather than alcohols, as we can observe that the number of water molecules is more than 15 times larger than that of ethanol molecules. From an initial inspection, one can see that both alcohol and water strong interacts with the carboxylic groups located at the edges of the GO sheets, which will contribute to further block the entrance of other molecules into the membrane. The easier molecular water permeation is favored due to its smaller size in relation to alcohol molecules. In principle, the membrane would work as a molecular sieving and the size selectivity would be the first explanation for the water/alcohol separation. However, when the porous channel width is increased to 10 Å, both water and alcohol molecules can easily get into the membrane at the initial stages, as shown in Figure S1. Although the separation is more effective for the cases of narrower channels, the favorable permeability of water can be also observed for the cases of GO10. This suggests that the size selectivity is not the unique relevant factor for the separation process.

In order to gain further insights on these features Grand-canonical Monte Carlo (GCMC) simulations to study the water/alcohol co-adsorption were performed. Figure 3 shows the ethanol/water and methanol/water co-adsorption isotherms at T=350K and mole fraction 0.1, 0.3 and 0.5 in the membrane GO7 (35%). At the initial adsorption stages, *i.e.*, at low pressure values, the amount of adsorbed alcohol and water are similar, while at high pressure (>0.1 bar) an effective separation was observed in all studied cases. The separation factor is calculated as $\alpha = (y_{water}/y_{alcohol})/(x_{water}/x_{alcohol})$, where *y* is the mole fractions in the adsorbed phase and *x* is the mole fractions in the solution, both at equilibrium. GO7 (35%) is the most effective selective membrane with $\alpha$ equal to 18.1 and 4.5 for water/ethanol and water/methanol, respectively. GCMC simulations indicate that the decreasing order of guest/host affinity is ethanol/GO, methanol/GO and water/GO, with adsorption enthalpy varying around -77, -73 and -64 kJ/mol, respectively (see Table S1). This affinity order is not what we could expect to explain the preferable adsorption of water inside these membranes. Indeed, the preferred sites of adsorption inside the membrane are very similar for both alcohols and water. They mainly interact with the functional groups (*e.g.* OH, COOH and O-epoxy) via H-bonds, as indicated in the hydrogen-oxygen, $g_{HcaO*}(r)$ and $g_{HhxO*}(r)$, pair distribution function displayed in Figure 4. The first adsorption site is not affected neither by the interlayer distance of the membrane nor by the functionalization level. The alcohol has additional affinity via hydrophobic interaction of $CH_3$ and $CH_2$ with pristine graphene region, which could explain the higher adsorption enthalpy values compared to water.

We have also investigated how the ratio of functional groups present in the GO structure affects the selectivity. When the GO O/C ratio was decreased from 35% to 14% the water/ethanol separation factor decreases from 18.1 to 5.4 for GO7 and from 3.2 to 2.3 for GO10. The water/methanol separation factor was substantially smaller than water/ethanol in all cases. Decreasing O/C ratio, the water/methanol separation factor decreases from 4.5 to 2.0 for GO7 and from 3.2 to 2.8 for GO10 (see Table S2). These selectivity values are quantitatively comparable and sometimes larger them those of porous materials used for alcohol/water separation[16,17]. Clearly, the membrane selectivity is enhanced by the functionalization degree, mainly in GO7 membranes. Although the addition of oxygen-containing functional groups increases the interaction sites for both water and alcohol molecules, it could be an additional constrain for the alcohol permeation. On the other hand, when decreasing the O/C ratio, the pristine regions are necessarily increased, then, the membrane becomes more hydrophobic and repulsive to water while being more attractive to alcohol. Then, to explain the preferred water adsorption, the preferable affinity of ethanol must be counter balanced by other effects such as steric hindrance. In other words, the separation phenomenon is not governed by affinities with the membrane (enthalpic mechanism) but rather by the geometry and size factors (entropic mechanism).

To confirm this hypothesis a careful analysis of the preferential arrangements for each component was performed. Figure 5 shows a snapshot of adsorbed molecules inside a single channel. The 2D pore geometry imposes the formation of a monolayer of adsorbed molecules. At this configuration, water can easily form 2D H-bonds network with up to 4 H-bonds per water molecule. This configuration is the most energetically favorable. Once the ethanol molecule is adsorbed it would disrupt this network. Ethanol can also form H-bonds via the OH groups but it has limited possibilities for that. The asymmetry of the alcohol molecule additionally to its geometry and large size make its alignment into the monolayer structure more difficult. Moreover, the ethanol has OH group being hydrophilic, while the rest of the ethanol, the $C_2H_5$ group, being repulsive to water. This leads to a natural phase separation inside the channels forming alcohol-rich and water-rich islands, similarly to what was observed in experimental analyses.[18] In other words, the adsorption selectivity mechanism is explained by the separation into a water phase, which is more favorable to growth inside the channel *vs.* an ethanol phase, which is less energetically favorable. Similar mechanism occurs with methanol/water separation. However, since methanol is smaller it has less constraining, the methanol phase is less unfavorable. These results validate the network model proposed by Geim *et al.*[4] to explain the GO selectivity.

Another important factor that influences the membrane selectivity is the guest diffusion inside the channels. Our results show that the water permeation is drastically decreased in presence of alcohol. The flow rates of pure water crossing the membrane are 23.16 and 247.63 #$H_2O$/ns for GO7 and GO10 (35%), respectively (see Figure S2). When the alcohol is mixed with water, the flux dropped one order of magnitude (see Table 1). When O/C ratio is reduced to 14% GO becomes more hydrophobic and it repeals water, thus increasing its diffusion inside the membrane. Thus, this results that the total flux crossing the membrane is increased in all cases, except for the case of GO7(35%) where flux is compensated by the fact that the membrane is mainly filled by fast diffusion water molecules. The presence of alcohol strongly reduces the flow through the membrane and it is inversely proportional to the size of the alcohol molecules.

The decreasing of the water flux is the first evidence that alcohol strongly influences the diffusion of water. This might be associated with the solvent structural rearrangement dynamics. The 2D pore geometry (in particular at very narrow channels, such as in the CO7, where the molecules are restricted to form only a single molecular layer) forces the fluid diffusion to be essentially two-dimensional. In this condition, the water has intrinsically faster diffusion than alcohols due to the less steric hindrance effects and also to the fact that co-planar water molecules can easily slip inside the channels. The fast water permeation against lower alcohol permeation enhances the separation/selectivity mechanism. This effect is more evident in membranes with low functionalization levels. For instance, for the case GO7(14%) the membrane has low adsorption selectivity, however the number of water molecules that effectively crossed the membrane is significantly larger than the corresponding alcohol molecules (see Figure S3). These processes can be better visualized from the videos in the supplementary materials.

In summary, there exist a balance among a number of factors that influence the preferable GO water selectivity. Although GO membranes have more affinity to alcohols, they are mostly hydrophilic which also attracts water. The 2D confinement restricts the molecules to stay in layered configurations, which is highly unfavorable for alcohols, while it is strongly favorable for water diffusion. Also, the 2D channel geometry is crucial for an effective separation/selectivity mechanism, since it forces the formation of water monolayers H-bonded network. In other words, the pore geometry of this membrane is the crucial feature that assist water selectivity.

**CONCLUSION**

A series of molecular dynamics and Grand-Canonical Monte Carlo simulations were performed to investigate the ethanol/water and methanol/water separation/selectivity through GO membranes with different interlayer distances and levels of functional groups. To rationalize the preferred selectivity of water in nano-confined structures and the effective blocking of alcohol molecules, we have performed a deep analysis on the structural molecular arrangements and diffusion mechanisms. We have concluded that both the size exclusion and molecular arrangements within 2D GO channels are responsible for the separation. The probable formation of water monolayer is enhanced by the formation of a robust 2D H-bond network. Moreover, the 2D channels favors small co-planar molecules such water to fast diffusion. The role of the number of functional groups in the membranes was also investigated and we observed that it contributes to increase the separation factor. A balance among diffusivity, membrane affinity, molecular size exclusion and geometry confinement are present during the separation mechanism. The presence of narrow 2D channels with high functionalization degree, such as in the case of GO7 (35%), appears to be the best membrane for ethanol/water separation. In other words, the separation phenomenon is not governed by affinities with the membrane (enthalpic mechanisms) but rather by the geometry and size factors (entropic mechanisms).

**METHODS**

Molecular dynamics simulations (MD) were performed using the structural model schematically shown in Figure 1. Typical structural models contain four graphene sheets perforated with nanoslits of width $D \sim 20$ Å and parallelly arranged by a distance $d$. We used the reflector wall protocol, where immaterial walls are placed on the extremity of the simulation box and parallel to the membrane to create a molecular flow through a unique path, as schematically shown in the inset of Figure 1. Two $d$ different values were considered here: 7 and 10 Å. The simulation box dimensions are 76.5 x 44.2 x 200 Å$^3$. Two types of graphene sheets were considered: graphene oxide (GO) with 35% and 14% content of oxygen (from hydroxyl, epoxy and carboxyl groups) atoms, as shown in Figure 1b. To build GO sheets, graphene membranes were functionalized with hydroxyl and epoxy functional groups on the both membrane sides and with carboxyl groups on the sheet edges.

The system configuration was then prepared by placing a water-alcohol liquid reservoir into contact to the fixed membrane. The initial water-ethanol and water-methanol configuration was generated using the Packmol[19] code and equilibrated at ambient pressure (1 atm) and room temperature (300 K) through MD simulations. Once the reservoir is placed into contact to the membrane, the water flow was simulated through controlling the reservoir thermodynamics properties. The reservoir temperature is kept constant using a Nosé-Hoover thermostat[20,21] and the pressure is controlled by a movable piston of graphene placed on the bottom of the simulation box. The piston allows the reservoir volume to vary while the water moves into the membrane. The piston position is scaled by the force experienced by the piston as a response of the reaction mixture force. This protocol is very effective to mimic an infinite reservoir.

The positions of the atoms of the membrane are kept fixed, so the bonded interaction description for the membrane is not needed. The Lennard-Jones parameters were extracted from CHARMM[22,23] force field for GO and the Lorentz-Berthelot mixing rules are used to determine the parameters for the cross-interactions[24,25]. Partial charge values were taken from Ref. [26]. The rigid extended simple point charge (SPC/E) model[27] was used to describe water molecules, while the TraPPE force field was used for ethanol and methanol[28]. The van der Waals interactions are truncated at 12 Å, and the long-range Coulomb interactions are computed by utilizing the particle–particle particle-mesh (PPPM) algorithm[29]. Periodic boundary conditions were imposed along the *xy*-plane containing the membranes, while non-periodic boundaries were used along the *z*-direction. The MD simulations were carried out using the open source software called large-scale parallel molecular dynamics simulation code (LAMMPS)[30].

Grand-Canonical Monte Carlo simulations were carried out to probe the water/methanol and water/ethanol co-adsorption at 350K. Three mixtures with mole fraction equal 0.1, 0.3 and 0.5 were considered. The same fixed multilayer graphene membrane model and the guest/host interactions using the classical force field formalism were applied as used in the MD simulations. The Ewald summation was also used for calculating the electrostatic interactions while the short-range contributions were computed with a cut-off distance of 12 Å. The GCMC simulations used to estimate the co-adsorption isotherms and adsorption enthalpy were computed using the revised Widom's test particle method[31].

**ACKNOWLEDGMENT**

This work was supported in part by the Brazilian Agencies CAPES, CNPq and FAPESP. The authors also thank the Center for Computational Engineering and Sciences at Unicamp for financial support through the FAPESP/CEPID Grant # 2013/08293-7.

**FIGURES**

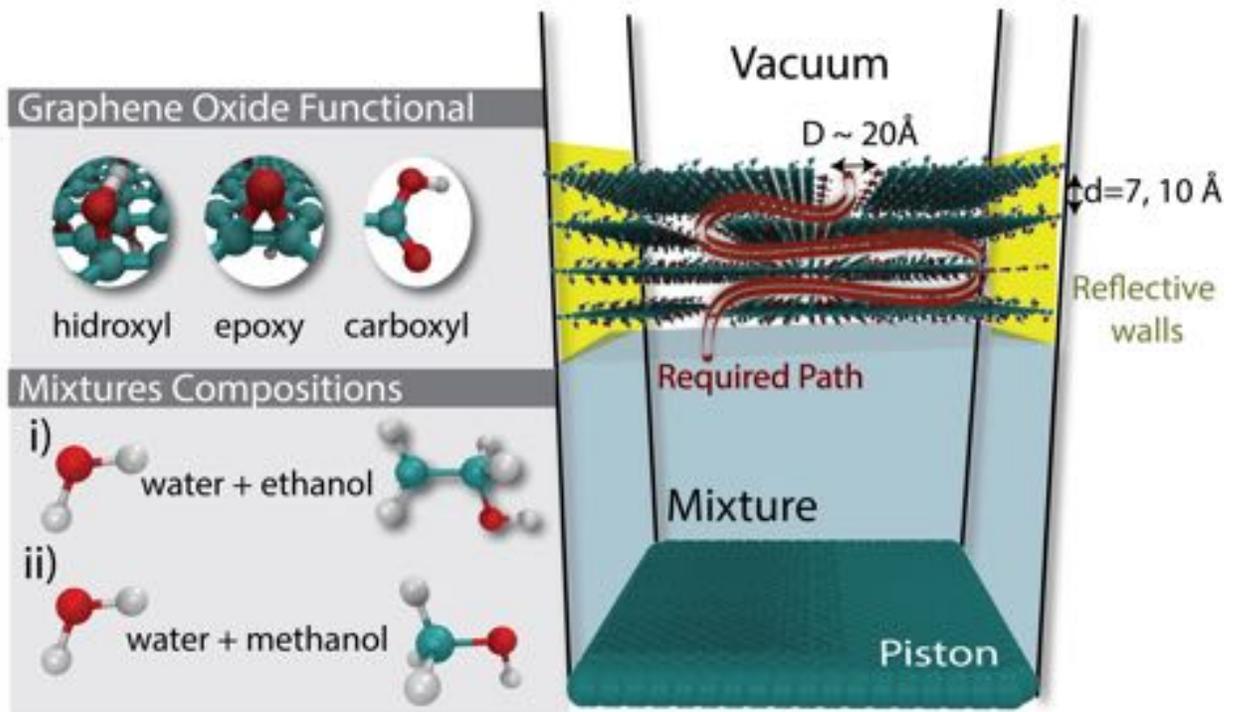

**Figure 1**. (Right) The simulated system composed of multilayer graphene-based sheets into contact with a mixture reservoir at constant temperature and pressure values, which are controlled by a movable piston at the bottom. Reflective walls (along the *yz*-plane) are placed in the box extremities to limit the mixture to flow through a predetermined pathway. (Left) The graphene oxide functional and mixture compositions are highlighted in the inset. Two types of graphene oxide sheet composed of 35% and 14% content of oxygen from hydroxyl, epoxy and carboxyl groups were considered. Two mixture compositions were considered: *i)* 50% water + 50% ethanol and; *ii)* 50% water and 50% methanol

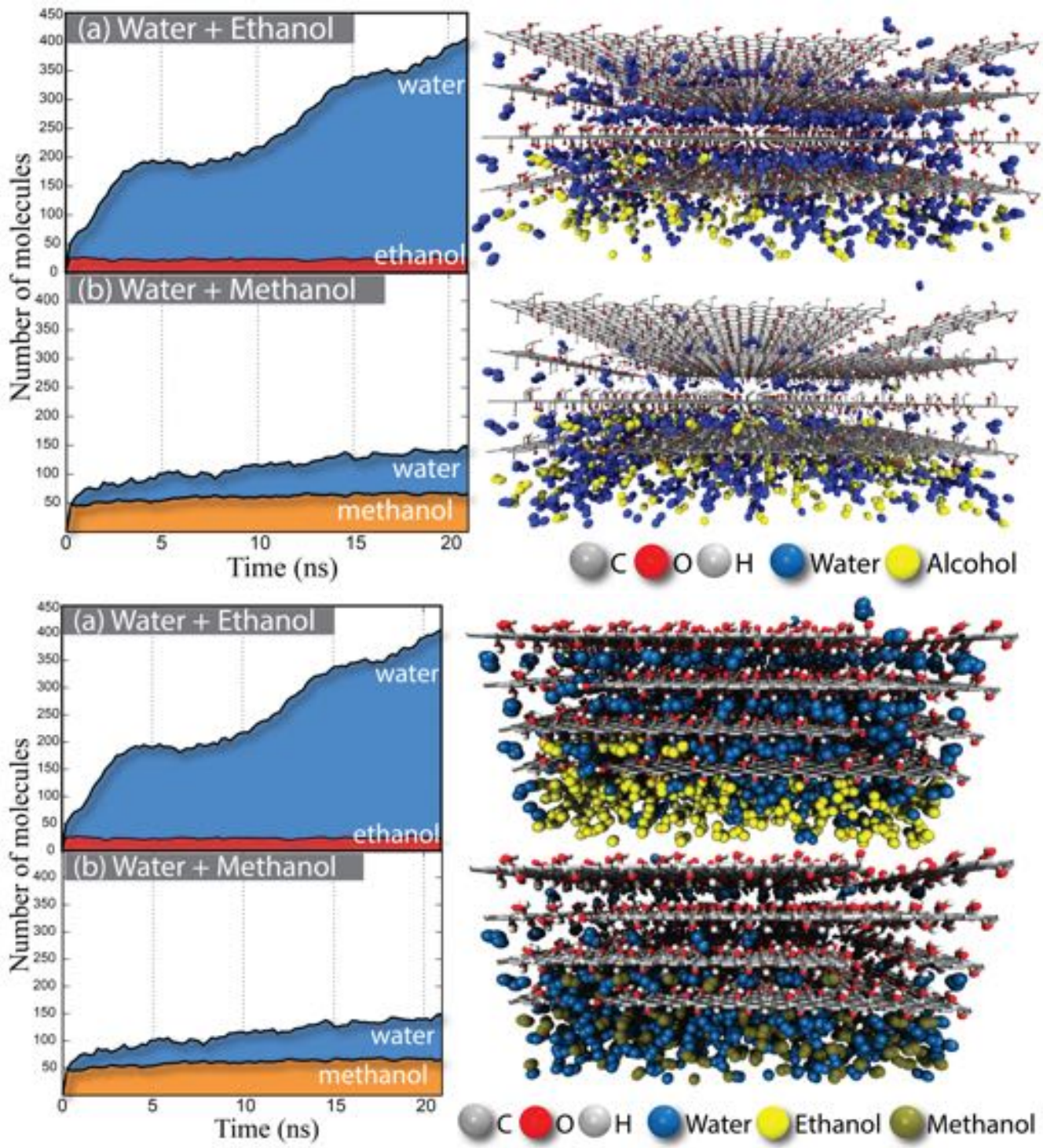

**Figure 2:** (Left) Time evolution of the number of molecules of ethanol/water (top) and methanol/water (bottom) permeating the GO7 for O/C equal to 35%. (Right) MD snapshots show the membrane being occupied mainly by water molecules (blue beads) while alcohols are rejected by the membrane.

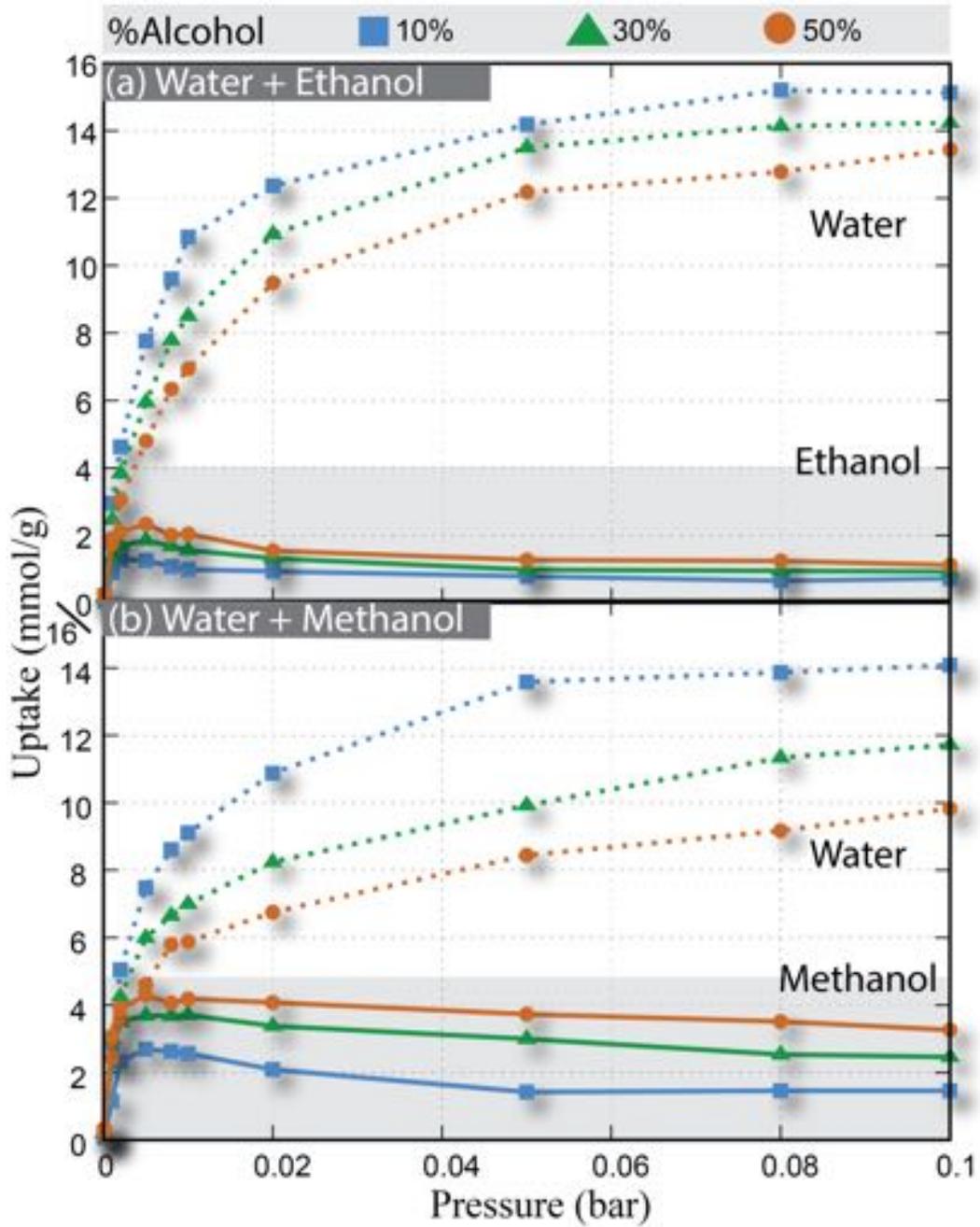

**Figure 3:** Grand-Canonical Monte Carlo results: (a) Ethanol/water and; (b) methanol/water co-adsorption isotherms

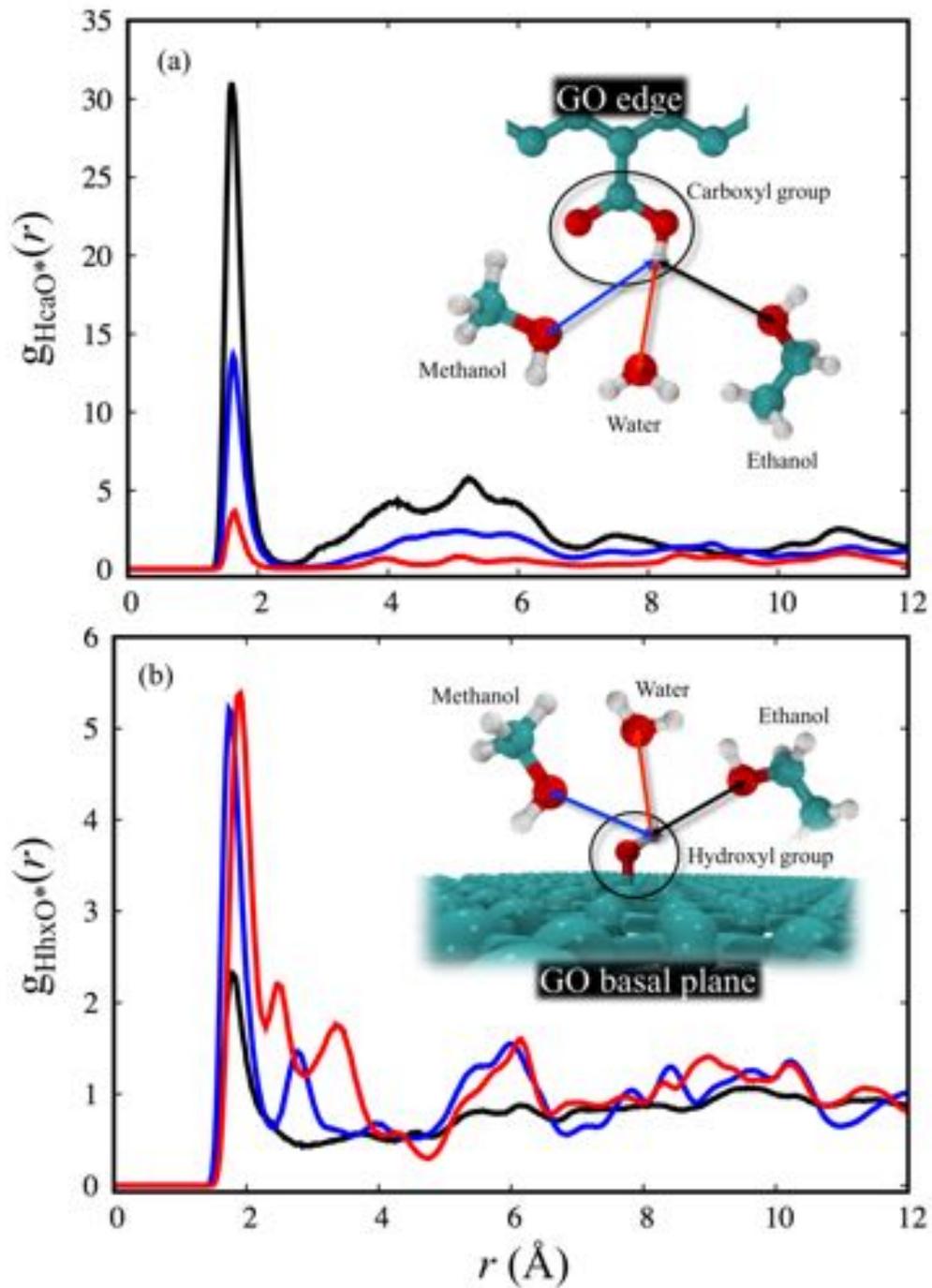

**Figure 4:** Radial distribution function between Oxygen atoms (*e.g.* Methanol - $O_{Me}$, Ethanol - $O_{Et}$ and Water - $O_w$ ) and Hydrogen atoms of (a) carboxyl (*Hca*), $g_{HcaO*}(r)$ and (b) hydroxyl (*Hhx*), $g_{HhxO*}(r)$. Water and alcohols have similar interactions with GO sheets via H-bonds with functional groups.

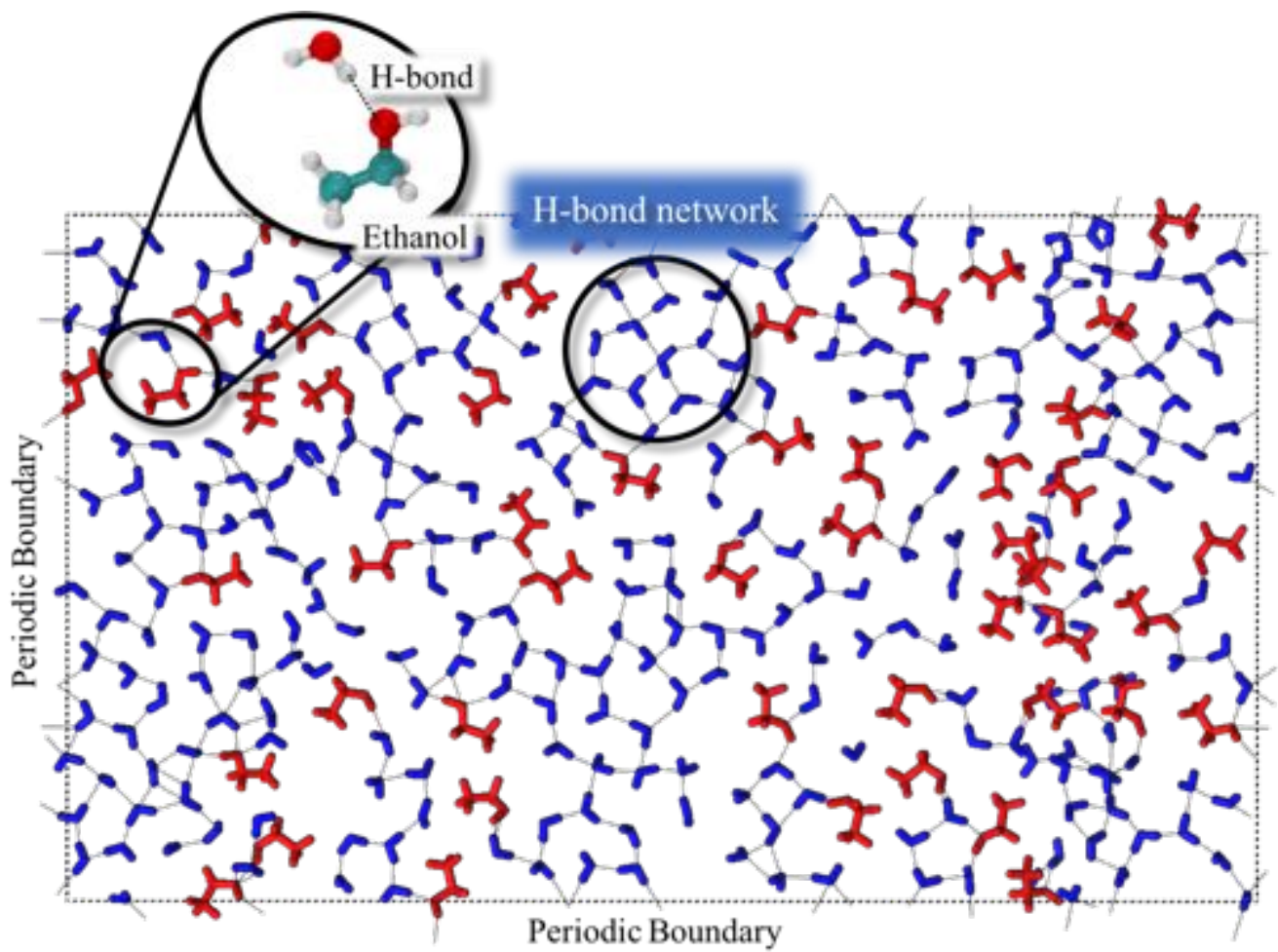

**Figure 5:** Adsorbate (*e.g.* water in blue and ethanol in red) monolayer between two GO sheets inside a GO7(35%) membrane. The coplanar water forms a 2D H-bond network while ethanol breaks it.

**Table 1:** Flux of mixture permeating the membrane

|  | Flux (#molecules/ns) | | | |
| --- | --- | --- | --- | --- |
| *Mixture* | *GO7(35%)* | *GO10(35%)* | *GO7(14%)* | *GO10(14%)* |
| Pure water | 23.16 | 247.63 | - | - |
| Water-ethanol | 2.16 | 22.43 | 1.01 | 66.05 |
| Water-methanol | 0.07 | 40.84 | 3.05 | 77.66 |

**REFERENCES**

[1] G. Liu, W. Jin, and N. Xu: Graphene-based membranes. *Chem. Soc. Rev.* **44**, 5016 (2015).

[2] Di An, Ling Yang, Ting-Jie Wang, and Boyang Liu. Separation Performance of Graphene Oxide Membrane in Aqueous Solution. *Ind. Eng. Chem. Res.,* **55 (17)**, 4803–4810 (2016)

[3] J. Abraham, K. S. Vasu, C. D. Williams, K. Gopinadhan, Y. Su, C. T. Cherian, J. Dix, E. Prestat, S. J. Haigh, I. V. Grigorieva, P. Carbone, A. K. Geim and R. R. Nair: Tunable sieving of ions using graphene oxide membranes. *Nature Nanotechnology* (2017) doi:10.1038/nnano.2017.21

[4] R. R. Nair, H. A. Wu, P. N. Jayaram, I. V. Grigorieva, and A. K. Geim: Unimpeded Permeation of Water Through Helium-Leak–Tight Graphene-Based Membranes. *Science* **335**, 442 (2012).

[5] R. K. Joshi, P. Carbone, F. C. Wang, V. G. Kravets, Y. Su, I. V. Grigorieva, H. A. Wu, A. K. Geim, and R. R. Nair. Precise and Ultrafast Molecular Sieving Through Graphene Oxide Membranes. *Science* **343**, 752 (2014).

[6] Y. Han, Z. Xu, C. Gao: Ultrathin Graphene Nanofiltration Membrane for Water Purification. *Adv. Funct. Mater.*, **23,** 3693–3700 (2013)

[7] K. Erickson, R. Erni, Z. Lee, N. Alem, W. Gannett, A. Zettl: Determination of the local chemical structure of graphene oxide and reduced graphene oxide. *Adv. Mater.* **22**, 4467-4472 (2010)

[8] J. Abraham, K. S. Vasu, C. D. Williams, K. Gopinadhan, Y. Su, C. T. Cherian, J. Dix, E. Prestat, S. J. Haigh, I. V. Grigorieva, P. Carbone, A. K. Geim and R. R. Nair: Tunable sieving of ions using graphene oxide membranes. *Nature Nanotechnology* (2017) doi:10.1038/nnano.2017.21

[9] C. Cheng, G. Jiang, C. J. Garvey, Y. Wang, G. P. Simon, J. Z. Liu, D. Li: Ion transport in complex layered graphene-based membranes with tuneable interlayer spacing. *Sci. Adv.* **2(2)**, 1501272 (2016)

[10] R. Luque, L. Herrero-Davila, J. M. Campelo, J. H. Clark, J. M. Hidalgo, D. Luna, J. M. Marinas and A. A. Romero: Biofuels: a technological perspective. *Energy Environ. Sci.*, **1**, 542–564 (2008)

[11] J. Hill, E. Nelson, D. Tilman, S. Polasky and D. Tiffany: Environmental, economic, and energetic costs and benefits of biodiesel and ethanol biofuels. PNAS, **103(30)**, 11206–11210 (2006)

[12] S. Gravelle, H. Yoshida, L. Joly, C. Ybert, and L. Bocquet: Carbon membranes for efficient water-ethanol separation. *The Journal of Chemical Physics* **145**, 124708 (2016)

[13] Z. Tian, S. M. Mahurin, S. Dai, and D. Jiang: Ion-Gated Gas Separation through Porous Graphene. Nano Lett., **17(3)**, 1802–1807 (2017)

[14] R. R. Nair, H. A. Wu1, P. N. Jayaram, I. V. Grigorieva, and A. K. Geim: Unimpeded Permeation of Water Through Helium-Leak–Tight Graphene-Based Membranes. *Science* **335**, 442 (2012).

[15] R. K. Joshi, P. Carbone, F. C. Wang, V. G. Kravets, Y. Su, I. V. Grigorieva, H. A. Wu, A. K. Geim, and R. R. Nair. Precise and Ultrafast Molecular Sieving Through Graphene Oxide Membranes. *Science* **343**, 752 (2014).

[16] A. Nalaparaju, X. S. Zhao and J. W. Jiang: Biofuel purification by pervaporation and vapor permeation in metal–organic frameworks: a computational study. *Energy Environ. Sci.*, **4**, 2107 (2011)

[17] J. Y. Lee, D. H. Olson, L. Pan, T. J. Emge, and J. Li: Microporous Metal–Organic Frameworks with High Gas Sorption and Separation Capacity, *Adv. Funct. Mater.*, 1**7**, 1255–1262 (2007)

[18] P. Bampoulis, J. P. Witteveen, E. S. Kooij, D. Lohse, B. Poelsema, and H. J. W. Zandvliet, Structure and Dynamics of Confined Alcohol–Water Mixtures, *ACS Nano*, **10 (7)**, pp 6762–6768 (2016)

[19] L. Martínez, R. Andrade, E. G. Birgin, and J. M. Martínez: PACKMOL: A package for building initial configurations for molecular dynamics simulations. *J. Comp. Chem.* **30**, 2157 (2009).

[20] S. Nosé: A unified formulation of the constant temperature molecular dynamics methods. *J. Chem. Phys.* **81**, 511 (1984).

[21] W. G. Hoover: Canonical dynamics: Equilibrium phase-space distributions. *Phys. Rev. A* **31**, 1695 (1985).

[22] K. Vanommeslaeghe, E. Hatcher, C. Acharya, S. Kundu, S. Zhong, J. Shim, E. Darian, O. Guvench, P. Lopes, I. Vorobyov, and A. D. MacKerell Jr.: CHARMM general force field: A force field for drug-like molecules compatible with the CHARMM all-atom additive biological force fields. *J. Comp. Chem.* **31**, 671 (2010).

[23] W. Yu, X. He, K. Vanommeslaeghe, and A. D. MacKerell Jr.: Extension of the CHARMM General Force Field to sulfonyl-containing compounds and its utility in biomolecular simulations. *J. Comp. Chem.*, **33**, 2451 (2012).

[24] H. A. Lorentz: On the application of the theorem in the kinetic theory of the gases, *Annalen der Physik,* **248**, 127-136 (1881).

[25] D. Berthelot : Gas Mixture, *Comptes rendus hebdomadaires des séances de l'Académie des Sciences*, **126**, 1703-1855 (1898).

[26] S. Jiao, and Z. Xu: Selective gas diffusion in graphene oxides membranes: a molecular dynamics simulations study. *ACS Appl. Mater. Interfaces*, **7**, 9052 (2015).

[27] H. J. C. Berendsen, J. R. Grigera, and T. P. Straatsma: The Missing Term in Effective Pair Potentials. *J. Phys. Chem.* **91**, 6269 (1987).

[28] B. Chen, J.J. Potoff, and J.I. Siepmann. Monte Carlo calculations for alcohols and their mixtures with alkanes. Transferable potentials for phase equilibria. 5. United-atom description of primary, secondary and tertiary alcohols. *J. Phys. Chem. B*, **105**, 3093-3104 (2001).

[29] J. W. Eastwood, and R. W. Hockney: Shaping the force law in two-dimensional particle-mesh models. *J. Comp. Phys.* **16(4)**, 342-359 (1974).

[30] S. Plimpton, Fast Parallel Algorithms for Short-Range Molecular Dynamics. *J. Comput. Phys.*, **117**, 1–19 (1995)

[31] B. Widom, Some Topics in the Theory of Fluids, *J. Chem. Phys.,* **39**, 2808 (1963).